\documentclass[draft,aps,pre,twocolumn,groupaddress,showkeys,nofootinbib,preprintnumbers,floatfix]{revtex4-2}
\usepackage{probs}
\usepackage{dynlearn}
\usepackage[linesnumbered,ruled]{algorithm2e}
\mathchardef\mhyphen="2D

\begin{document}

\def\ourTitle{Exploring Predictive States via Cantor Embeddings and Wasserstein Distance
}

\def\ourAbstract{Predictive states for stochastic processes are a nonparametric and interpretable
construct with relevance across a multitude of modeling paradigms. Recent
progress on the self-supervised reconstruction of predictive states from
time-series data focused on the use of reproducing kernel Hilbert spaces. Here,
we examine how Wasserstein distances may be used to detect predictive
equivalences in symbolic data. We compute Wasserstein distances between
distributions over sequences (``predictions''), using a finite-dimensional
embedding of sequences based on the Cantor for the underlying geometry. We show
that exploratory data analysis using the resulting geometry via hierarchical
clustering and dimension reduction provides insight into the temporal structure
of processes ranging from the relatively simple (\emph{e.g.}, finite-state
hidden Markov models) to the very complex (\emph{e.g.}, infinite-state indexed
grammars).
}

\def\ourKeywords{time series, predictive states, Wasserstein distance, hierarchical clustering
}

\hypersetup{
  pdfauthor={Samuel Loomis},
  pdftitle={\ourTitle},
  pdfsubject={\ourAbstract},
  pdfkeywords={\ourKeywords},
  pdfproducer={},
  pdfcreator={}
}

\author{Samuel P. Loomis}
\email{sloomis@ucdavis.edu}

\author{James P. Crutchfield}
\email{chaos@ucdavis.edu}
\affiliation{Complexity Sciences Center and Department of Physics and Astronomy,
University of California at Davis, One Shields Avenue, Davis, CA 95616}

\date{\today}
\bibliographystyle{unsrt}

\title{\ourTitle}

\begin{abstract}
\ourAbstract
\end{abstract}

\keywords{\ourKeywords}

\preprint{\arxiv{2206.XXXXX}}

\date{\today}
\maketitle

\makeatletter
\let\toc@pre\relax
\let\toc@post\relax
\makeatother

\setstretch{1.1}

\listoffixmes

\section{Introduction}
Suppose that we have a finite sequence $x_1 \dots x_L$ of categorical
observations drawn from a temporal process. We may suppose that the process is
stationary (time-translation invariant) and ergodic (explores
all possible behaviors) \cite{Fell70a,Kall01a}. We may wish to forecast from the
observed information the behavior of the next $n$ observations. If we suspect
that the process' temporal correlations do not matter much beyond $k$ symbols,
then we take our data to be all subsequences of length $n+k$,
splitting each subsequence into words of length $k$ and $n$ respectively. From
these ``past/future'' pairs we can construct an empirical conditional
distribution
\begin{align*}
  \widehat{P}_{x_{1}\dots x_{n}|x_{-k+1}\dots x_0} = 
  \frac{C_{x_{-k+1}\dots x_{n}}}{C_{x_{-k+1}\dots x_0}}
  ~,
\end{align*}
where $C_{w}$ is the number of times the word $w$ appears in our sequence $x_1
\dots x_L$. 

Suppose, though, that we do not know how long-range the process' temporal
dependencies can stretch. Even very simple stochastic processes can have
infinite Markov order, indicating potential long-term dependence of future
observations on the past \cite{Uppe97a}. Given sufficient data, it would be
desirable to take pasts of arbitrary length and converge towards a prediction
conditioned on the \emph{infinite} past:
\begin{align}
  {P}_{x_{1}\dots x_{n}|\overleftarrow{x}}
  = \lim_{k\rightarrow\infty}
  \widehat{P}_{x_{1}\dots x_{n}|x_{-k}\dots x_0}
\label{eq:convergence}
\end{align}
with the infinite sequence $\overleftarrow{x} = (\dots,x_{-1},x_0)$ of
observations stretching into the past. This mathematical ideal is known as the
\emph{causal} or \emph{predictive state} \cite{Crut88a,Jaeg00a}. Formally, the
conditional predictions ${P}_{x_{1}\dots x_{n}|\overleftarrow{x}}$ for all
forecast lengths $n$ together describe a \emph{probability measure} over future
sequences $\overrightarrow{x} = (x_1,x_2,\dots)$; the predictive state is this
measure. We denote it simply $P_{\overleftarrow{x}}$.

Predictive states are employed for inference and modeling in dynamical systems
\cite{Crut92c}, renewal processes and neural spike-trains
\cite{Marz14e,Marz15a}, condensed matter physics \cite{Varn14a}, and
spatiotemporal systems \cite{Rupe19b}. A deep mathematical
theory of predictive-state inference has been correspondingly developed
\cite{Uppe97a,Jaeg00a,Shal02a,Jame04a,Stil07b,Stre13a,Thon15a,Brod20a}.
If the dataset $x_1 \dots x_L$ is drawn from any stationary process and if $L$
is sufficiently large, then Eq. \eqref{eq:convergence} converges for any $n$
and a probability-$1$ subset of pasts $\overleftarrow{x}$ \cite{Uppe97a}.
Recently, this result has been further clarified: in the language of measures,
$P_{\overleftarrow{x}}$ converges \emph{in distribution}, with respect to the
\emph{product topology} of the space of sequences $\mathcal{X}^\mathbb{N}$
\cite{Loom21a}.

The broad goals of predictive-state analysis are threefold \cite{Shal98a}. The
first is to understand the overall structure of how the predictive states
relate to one another geometrically and, possibly, use this geometry to
classify pasts based on equivalence of their predictive states. The second is
to actually reproduce the prediction to a specified accuracy. The third is to
understand the dynamics of how predictions evolve under a stream of new
observations. 

The following focuses on the first, as it is a crucial building block to
achieving the other two. Recent attempts to reconstruct the geometry of
predictive states embedded them in reproducing kernel Hilbert spaces
\cite{Song09a,Song10a,Boot13a,Brod20a,Loom21a}. This was achieved to great
effect largely since convergence in Hilbert spaces generated by universal
kernels is equivalent to convergence in distribution \cite{Srip10a}. That is,
these embeddings allow accurate representations of predictive states because
they respect the product topology of sequences in the same way that predictive
states themselves do \cite{Loom21a}. Here, we achieve the same results using a
procedure that can be easily visualized and interpreted. This makes it suitable
as a method for exploratory data analysis \cite{Tuke62a,Tuke77a}, while moving
further toward interpretable machine learning of structured processes.

Understanding the source of the power of RKHS methods in predictive-state
analysis frees us to consider other options. The following embeds
symbolic sequences in a one-dimensional space that has the same topology as the
product space. This embedding is inspired by the fractal Cantor set
\cite{Kurk03a}. Predictive states can then be thought of as distributions in
this one-dimensional space. We then use the Wasserstein distance to compute the
geometry between predictive states, which is determined by a closed-form
integral for one-dimensional distributions. This operates as an alternative to
RKHS-based distances since the Wasserstein distance also reproduces the
topology of convergence in distribution \cite{Pane19a}. The resulting distance
matrix then is used to find low-dimensional embeddings \cite{Borg05a} of the
geometry or hierarchical clusterings \cite{Mull11a} of the predictive states.
When combined with the fractal embedding, the latter, in particular, provides a
highly interpretable visualization of the predictive-state space.

\section{Example processes}

The methods here are intended to be applied to stationary and ergodic
stochastic processes that generate categorical time-series data. For these
purposes we consider a stochastic process to be a collection of probability
distributions $\Prob{\mu}{x_1\dots x_L}$ over any finite, contiguous sequence,
taking values in a finite set $\mathcal{X}$. Formally, this describes a measure
$\mu$ over the set of all bi-infinite sequences $(\dots,x_{-1},x_0,x_1,\dots)
\in \mathcal{X}^{\mathbb{Z}}$.

These processes are generated by a number of systems with widely varying
complexity. Most popularly studied are those often characterized as having a
degree of ``finite memory'': \emph{Markov chains}, \emph{hidden Markov chains},
and \emph{observable operator models} (also termed \emph{generalized hidden
Markov chains}) \cite{Uppe97a,Jaeg00a}. Beyond these, one can also generate
processes using probabilistic grammars, such as probabilistic context-free and
indexed grammars \cite{Gema00a}. Additionally, coarse-grained data from chaotic
dynamical systems---such as the \emph{logistic map}---display behavior varying
widely in complexity \cite{Crut92c}.

We refer back frequently to the following example processes of increasing
computational complexity:
\begin{enumerate}
      \setlength{\topsep}{-2pt}
      \setlength{\itemsep}{-2pt}
      \setlength{\parsep}{-2pt}
\item The \emph{even process} can be generated by repeatedly tossing a coin
	and writing down a $0$ for every tail and $11$ for every head. The process
	is essentially random except that $1$s only appear in contiguous blocks of
	even size bounded by $0$s. The even process has infinite Markov order but
	can be generated by a two-state hidden Markov chain \cite{Crut01a}. A
	typical example might look like $01100111101100011$.
\item The $\mathtt{a}^n \mathtt{b}^n$ \emph{process} can be generated by
	choosing a random integer $n\geq 1$ (we suppose via a Poisson process) and
	writing $n$ $\mathtt{a}$s followed by an equal number of $\mathtt{b}$s,
	and then repeating this procedure indefinitely. This results in sequences
	where any contiguous block of $\mathtt{a}$s is followed by a block of
	$\mathtt{b}$s of equal size. The $\mathtt{a}^n \mathtt{b}^n$ process cannot
	be generated by any finite hidden Markov chain, though it is a simple
	example of a probabilistic context-free language \cite{Hopc06a}. A typical
	example might look like $\mathtt{abaaabbbabaabb}$.
\item The $x+f(x)$ \emph{process} is a probabilistic context-free language
	modeling the syntactic structure of simple mathematical expressions. It has
	terminal symbols $\{\mathtt{(}\ ,\ \mathtt{)}\ ,\ \mathtt{;}\ ,\
	\mathtt{+}\ ,\ \mathtt{f}\ ,\ \mathtt{x}\}$ and nonterminals $\{A,B,C\}$,
	and starts with a sequence of $A$s. Sequences are generated by applying the
	production rules:
\begin{align*}
    A &\mapsto B\ \mathtt{+}\ C\ \mathtt{;}\ |\ C\ \mathtt{;}\\
    B &\mapsto B\ \mathtt{+}\ C\ |\ C\\
    C &\mapsto \mathtt{f(}B\mathtt{)}\ |\ \mathtt{x}
  ~.
\end{align*}
	A typical example might look like $\mathtt{x+x;f(x+f(x));f(f(x));}$.
\item The $\mathtt{a}^n \mathtt{b}^n \mathtt{c}^n$ \emph{process} is a
	probabilistic indexed language \cite{Hopc06a} that is analogous to
	$\mathtt{a}^n \mathtt{b}^n$ except after writing the blocks of
	$\mathtt{a}$'s and $\mathtt{b}$'s, we also write a block of $\mathtt{c}$'s
	of length $n$. A typical example might look like
	$\mathtt{abcaaabbbcccabcaabbcc}$.
\item The \emph{Morse-Thue process} is generated by sampling from the time
	series of the logistic map at critical ``onset of chaos'' parameter
	$r_c \approx 3.56995$:
\begin{align*}
    y_{t+1} = r y_t(1-y_t)
\end{align*}
	and then coarse-graining the data by taking $x_t = 0$ if $0< y_t \leq
	\frac{1}{2}$ and $x_t = 1$ if $\frac{1}{2} < y_t <1$ \cite{Kurk03a}.
	Alternatively, we can generate this process by starting with a single $0$
	and executing the replacements $0 \mapsto 11$ and $1 \mapsto 01$
	consecutively. The resulting process is an indexed-context free language
	\cite{Crut92c}. A typical example might look like
	$11011101010111011101110101011101$---the fifth generation of the
	replacement rule starting from $0$.
\end{enumerate}

\begin{figure*}[ht]
\centering
\includegraphics[width=0.46\linewidth]{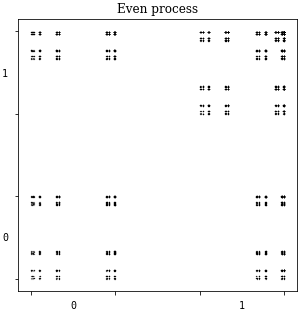}
\includegraphics[width=0.45\linewidth]{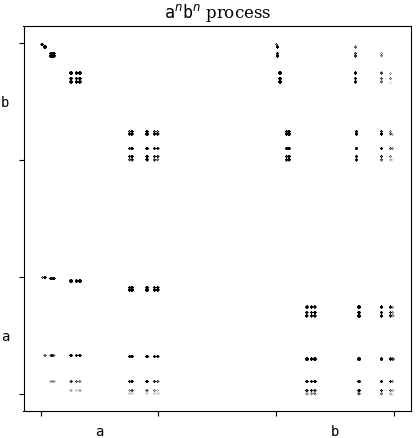}
\includegraphics[width=0.45\linewidth]{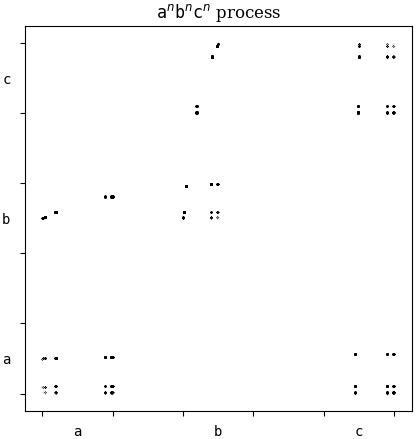}
\includegraphics[width=0.45\linewidth]{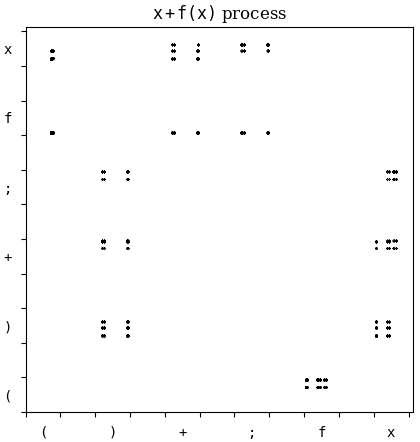}
\caption{Cantor plots for the even, $\mathtt{a}^n\mathtt{b}^n$,
	$\mathtt{a}^n\mathtt{b}^n\mathtt{c}^n$, and $\mathtt{x+f(x)}$ processes.
	Each point $(x,y)$ corresponds to a pair of sequences corresponding to the
	past $x$ and future $y$, respectively. The symbol on the $x$ ($y$) axis
	indicates that all points above (to the right of) that symbol have a past
	(future) whose most recent observation is that symbol. Though not marked,
	further proportional subdivisions of each segment of the axes indicate the
	value of the second, third, and so on symbols. For instance, one can read
	from the $\mathtt{x+f(x)}$ fractal that any past ending in $\mathtt{f}$
	must be paired with a future beginning in $\mathtt{(f}$ or $\mathtt{(x}$.
	}
\label{fig:cantor}
\end{figure*}

\section{Cantor-embedding sequences}

The geometry of sequences is inherently self-similar. Given an infinite sequence
$\overrightarrow{x} = (x_1,x_2,\dots)$, we can split it into its leading word
$x_1 x_2 \dots x_L$ and a following sequence $\overrightarrow{x}_L =
(x_L,x_{L+1},\dots)$. That is, the space of sequences $\mathcal{X}^\mathbb{N}$
can be factored into $\mathcal{X}^L\times \mathcal{X}^\mathbb{N}$ for any $L$.
The fractal nature of sequence-space is encoded in the structure of its
\emph{product topology}.

We exploit this self-similarity in an interesting way by constructing a mapping
between sequence space and the celebrated \emph{Cantor set} (or one of its
generalizations). Suppose a symbolic sequence $(x_1,x_2,\dots)$ takes values in
an alphabet $\mathcal{X}$ of size $|\mathcal{X}|$. To each $x\in\mathcal{X}$ we
associate a unique integer between $0$ and $|\mathcal{X}|-1$ inclusive; call
this $J(x)$. Then, there is a function $C:\mathcal{X}^\mathbb{N}\rightarrow
[0,1]$ that maps every sequence to a positive real number:
\begin{align*}
  C(x_1,x_2,\dots) = \sum_{k=1}^\infty \frac{2J(x_k)}{(2|\mathcal{X}|-1)^k}
  ~.
\end{align*}
For instance, suppose that $|\mathcal{X}|=2$ has two elements; then $C$ maps the
sequence to a point the traditional Cantor set fractal. For a finite sequence of
length $L$, truncate the sum at $k=L$.

Remarkably, the embedding $C$ has the property that for any continuous function
$f$ on $[0,1]$, the function $F(\overrightarrow{x}) = f(C(\overrightarrow{x}))$
is continuous on $\mathcal{X}^\mathbb{N}$. Further, if $F$ is continuous on
$\mathcal{X}^\mathbb{N}$, then $f(y) = F(C^{-1}(y))$ is continuous on the image.
Thus, the embedding $C$ respects the basic structure of the product topology
\cite{Kurk03a}.

Stationary processes, due to their time-translation invariance, inherit the
fractal temporality of sequence space. This can be easily visualized: Given a
length-$L$ sample $x_1\dots x_L$, and $n,k>0$, take a sliding window of pasts
and futures, $(x_{t-k+1}\dots x_t, x_{t+1}\dots x_n)$ for $t=k,\dots,L-n$.  For
each past-future pair, compute the truncated Cantor embeddings on the
\emph{reversed} past and (unreversed) future: $\left(C(x_t \dots
x_{t-k+1}),C(x_{t+1}\dots x_n)\right)$. The resulting pairs of real numbers can
be plotted as $(x,y)$-values on a scatter plot. The fractal that emerges
contains, in essence, all information necessary to understand a process'
temporal structures. See Fig. \ref{fig:cantor} for examples and guidance on how
to interpret the visualization.

Note that for $|\mathcal{X}|>2$ the embedding $C$ introduces additional
structure that may or may not be desired. Associating each symbol $x$ with an
integer $j_x$ endows an ordinal structure on the set $\mathcal{X}$. This
ordinality is present in the macroscopic geometry of
$C\left(\mathcal{X}^{\mathbb{N}}\right)$. Later, we examine higher-dimension
embeddings that do not have this. However, they come at the cost of increased
computational complexity when determining the Wasserstein distance. So, for now,
we assume that ordinal artifacts are either desired or sufficiently tolerable
as to not outweigh the computational benefits of working in one dimension.
\begin{algorithm}[t]
  \SetKwData{UnqPasts}{UnqPasts}\SetKwData{Cantors}{Cantors}\SetKwData{Wass}{Wass}
  \SetKwFunction{Wasserstein}{Wasserstein}\SetKwFunction{Append}{append}
  \SetKwFunction{Index}{index} \SetKwFunction{Length}{length}
  \SetKwFunction{Matrix}{Matrix}
  \SetKwFunction{CantorWasserstein}{CantorWasserstein}
  \SetKwInOut{Input}{input}\SetKwInOut{Output}{output}
  
  \underline{function \CantorWasserstein} ($k,n,x_1\dots x_L$)\;
  \Input{Integers $k,n$ of past and future lengths} 
  \Input{Length-$L$ sequence $x_1\dots x_L$ of observations} 
  \Output{List $\UnqPasts$ of unique pasts}
  \Output{List of lists $\Cantors$ of Cantor-embedded futures} 
  \Output{Matrix $\Wass$ of Wasserstein distances} 
  $\UnqPasts \leftarrow \left[\right]$\; 
  $\Cantors \leftarrow \left[\right]$\; 
  \For{$t\leftarrow n$ \KwTo $L-k$} { 
    $\overleftarrow{\mathbf{x}} \leftarrow \left[x_{t-k+1},\dots, x_{t}\right]$\;
    $\overrightarrow{\mathbf{x}} \leftarrow \left[x_{t+1},\dots, x_{t+n}\right]$\;
      \vspace{-1.5em}
      \begin{flalign*}
        p &\leftarrow \sum_{\ell=1}^n
        \frac{2J(\overrightarrow{x}_\ell)}{(2|\mathcal{X}|-1)^\ell}\ ; &&
      \end{flalign*}

      \eIf{$\overleftarrow{\mathbf{x}} \in \UnqPasts$} { 
        $\Append\ \overleftarrow{\mathbf{x}}$ \KwTo $\UnqPasts$\; 
        $\Append\ [p]$ \KwTo $\Cantors$\; 
        } { 
        $j \leftarrow \Index\left(\overleftarrow{\mathbf{x}},\UnqPasts\right)$\;
        $\Append\ p$ \KwTo $\Cantors_j$\; 
        } 
  }
  $K \leftarrow \Length(\UnqPasts)$\; 
  $\Wass \leftarrow \Matrix(K,K)$\;
  \For{$i\leftarrow 1$ \KwTo $K$} { 
    \For{$j\leftarrow 1$ \KwTo $K$} {  
      $\Wass_{ij}
        \leftarrow
        \Wasserstein(\Cantors_i,\Cantors_j)$\;
        $\Wass_{j i} \leftarrow \Wass_{i j}$\; } } 
  \KwResult{$\UnqPasts$,$\Cantors$,$\Wass$}
\caption{Convert a sequence of categorical time-series data into a labeled
	collection of empirical distributions of Cantor-embedded futures and a
	matrix of Wasserstein distances between said distributions.
	}
\label{alg:cantorwass}
\end{algorithm}

\section{Wasserstein distance on predictive states}

Figure \ref{fig:cantor}'s Cantor fractals represent probability distributions:
We interpret a vertical slice of the fractal, located at horizontal position
$C(\overleftarrow{x})$, as visualizing the predictive state
$P_{\overleftarrow{x}}$ as a distribution over Cantor-embedded futures
$C(\overrightarrow{x})$.

For example, by examining the even process' Cantor fractal, one notices that
there are effectively only $2$ distinct predictive states---every vertical
column is just one of two types. This corresponds with the $2$ states of the
hidden Markov model that generates the even process.

This allows us to see how predictive states distribute their probability over
the intrinsic geometry of potential futures. We compare predictive states not
only on how much their supports overlap, but on how geometrically close their
supports are to one another. For the $\mathtt{a}^n \mathtt{b}^n$ process, for
example, we see that the first few columns (corresponding to pasts of the form
$\dots \mathtt{b a}^n$ for some $n$) are inherently similar to one another,
though they are shifted upwards the closer to the axis they are. (The latter
corresponds to the increasing number of $\mathtt{b}$s in the predicted future
as $n$ increases.)

The intuitive distance metric between probability measures for capturing this
underlying geometry is the Wasserstein metric \cite{Pane19a}. Given two measures
$\mu$ and $\nu$ defined on a metric space $\mathcal{M}$ with metric $d$, the
Wasserstein distance between $\mu$ and $\nu$ is given by:
\begin{align*}
  W(\mu,\nu) = \min_{\pi\in \Gamma(\mu,\nu)}
   \int_{\mathcal{M}\times\mathcal{M}} 
   d(x,y) d\pi(x,y)
   ~,
\end{align*}
where $\Gamma(\mu,\nu)$ is the set of all measures on $\mathcal{M} \times
\mathcal{M}$ whose left and right marginals are $\mu$ and $\nu$, respectively.
It is the minimal cost to ``shift'' the probability mass from one distribution
to match the other's shape. 

$W(\mu,\nu)$ is the solution to a constrained linear optimization. As a
function of distributions, $W(\mu,\nu)$ is continuous with respect to
convergence in distribution. In fact, convergence under the Wasserstein
distance is equivalent to convergence in distribution on compact spaces. This
makes $W(\mu,\nu)$ ideal for measuring geometry between predictive states,
since empirical estimates of these are known to converge in distribution.

When $\mathcal{M} \subseteq \mathbb{R}$, there is in fact a closed-form solution
to the Wasserstein optimization problem \cite{Thas10a}. Let $F$ and $G$
respectively be the cumulative distributions functions of $\mu$ and $\nu$.
Then:
\begin{align*}
  W(\mu,\nu) = \int_{-\infty}^\infty \left|F(t)-G(t)\right| dt
  ~.
\end{align*}
This closed-form solution is considerably faster to compute than the linear
optimization required for arbitrary metric spaces. Since the Cantor embedding
embeds the space of sequences directly into $[0,1]$, we can directly employ
this formula.

Combining the Cantor embedding and Wasserstein distance leads to a
straightforward program for analyzing categorical time series:
\begin{enumerate}
\item Apply Algorithm \ref{alg:cantorwass} to the data stream $x_1\dots x_L$
	for a specified past length $k$ and future length $n$, retrieving the (i)
	set of unique observed pasts, (ii) empirical Cantor distributions
	corresponding to each past, and (iii) matrix of Wasserstein distances
	computed from these distributions.
\item To elucidate the relative geometry of the predictive states, use the
	Wasserstein distance matrix to perform additional methods of geometric data
	analysis, such as hierarchical clustering \cite{Mull11a} and
	multidimensional scaling \cite{Borg05a}.
\end{enumerate}
The next two sections examine the results of this approach.

\begin{figure*}[ht]
\centering
\includegraphics[width=0.49\linewidth]{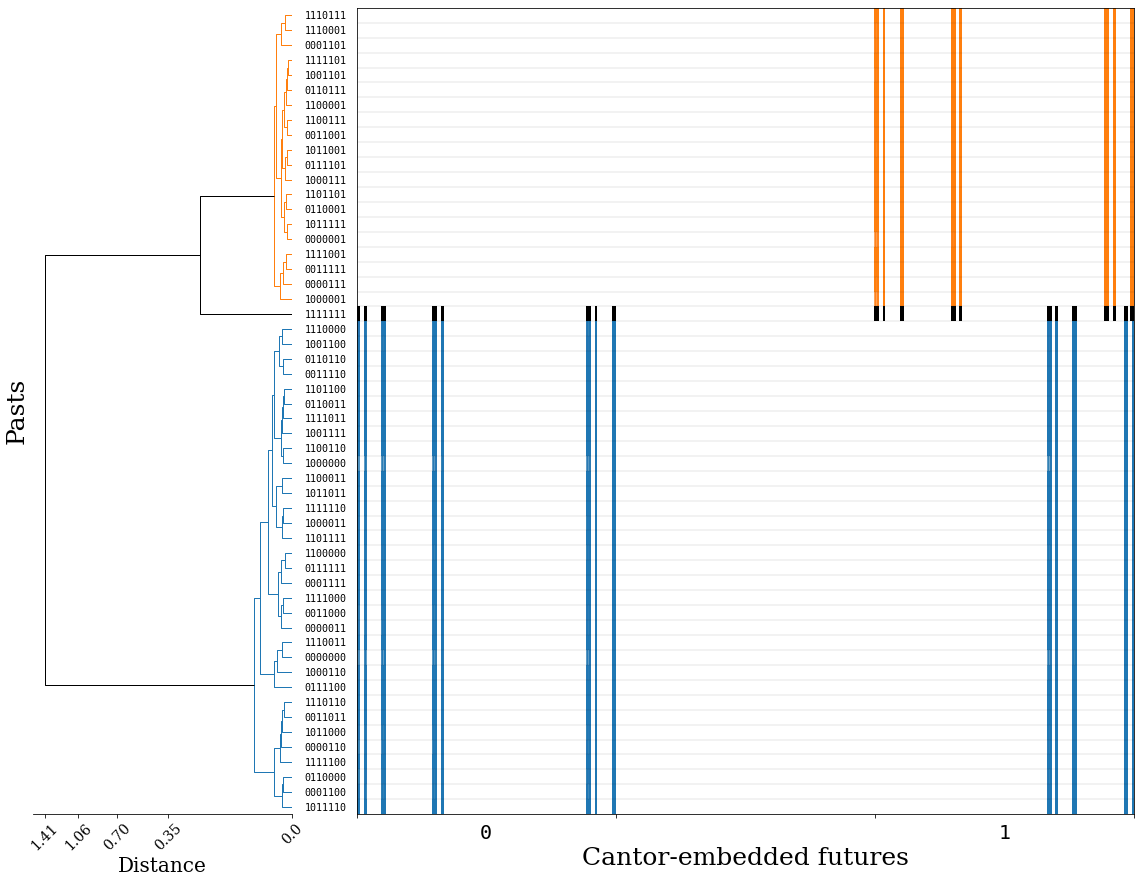}
\includegraphics[width=0.49\linewidth]{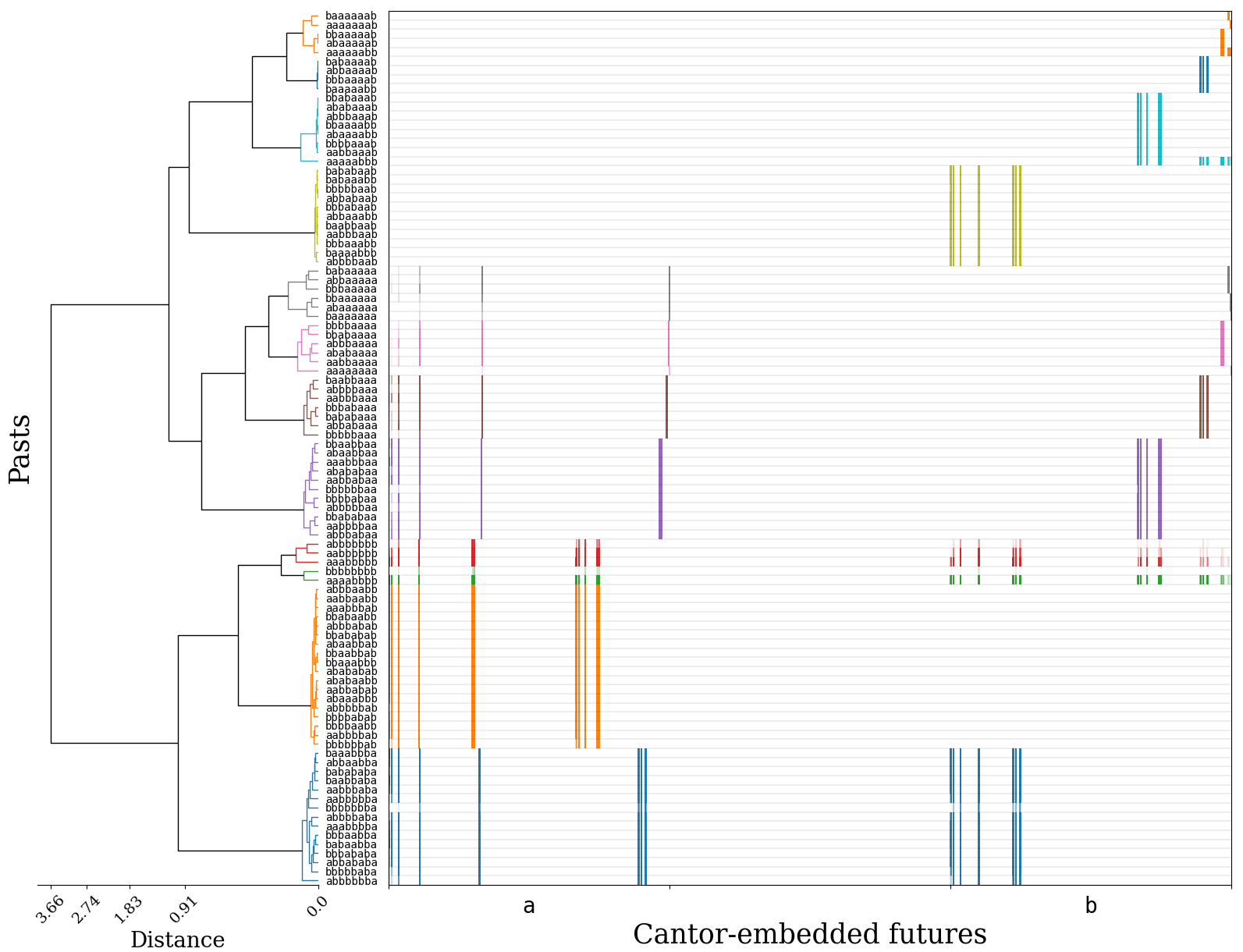}
\includegraphics[width=0.49\linewidth]{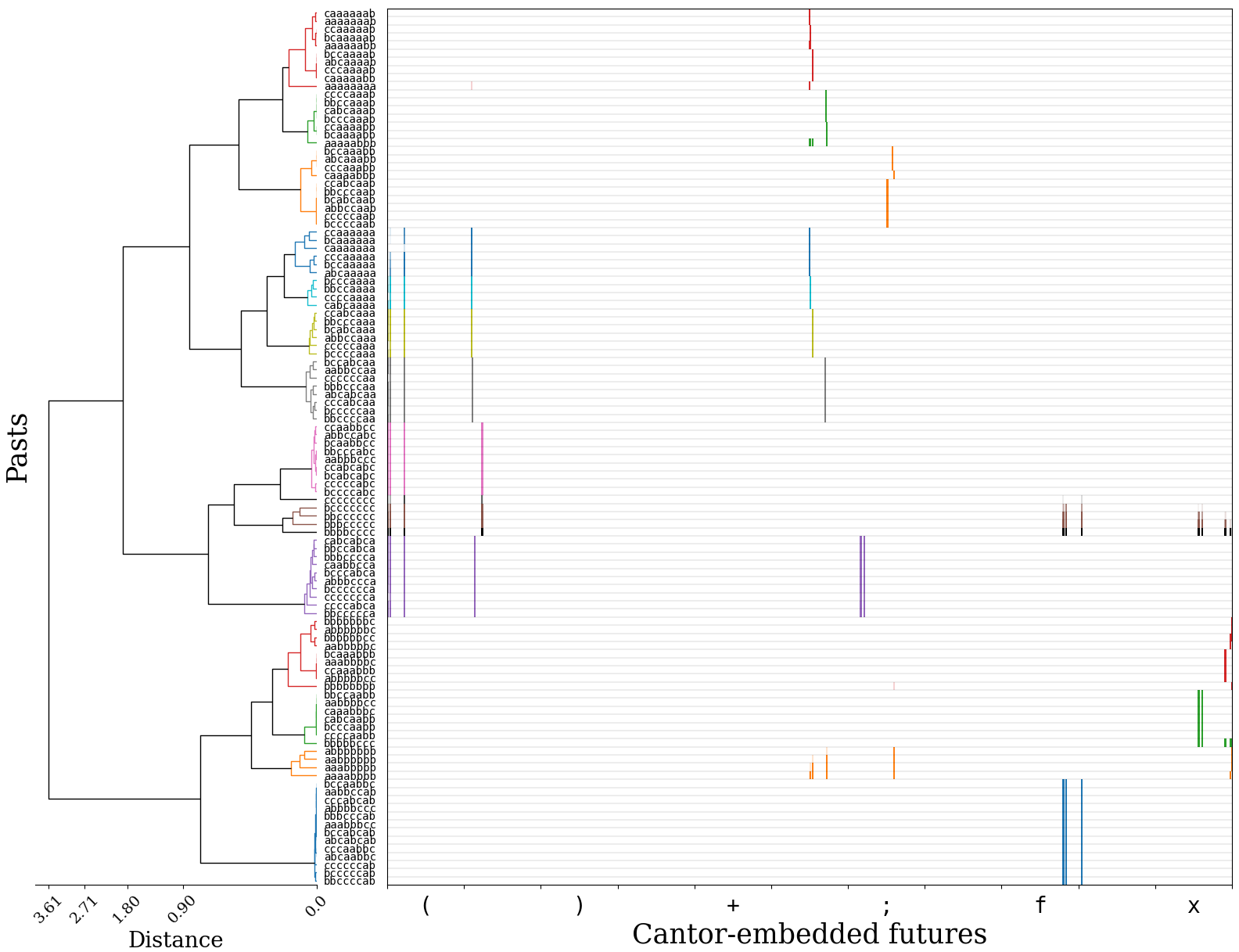}
\includegraphics[width=0.49\linewidth]{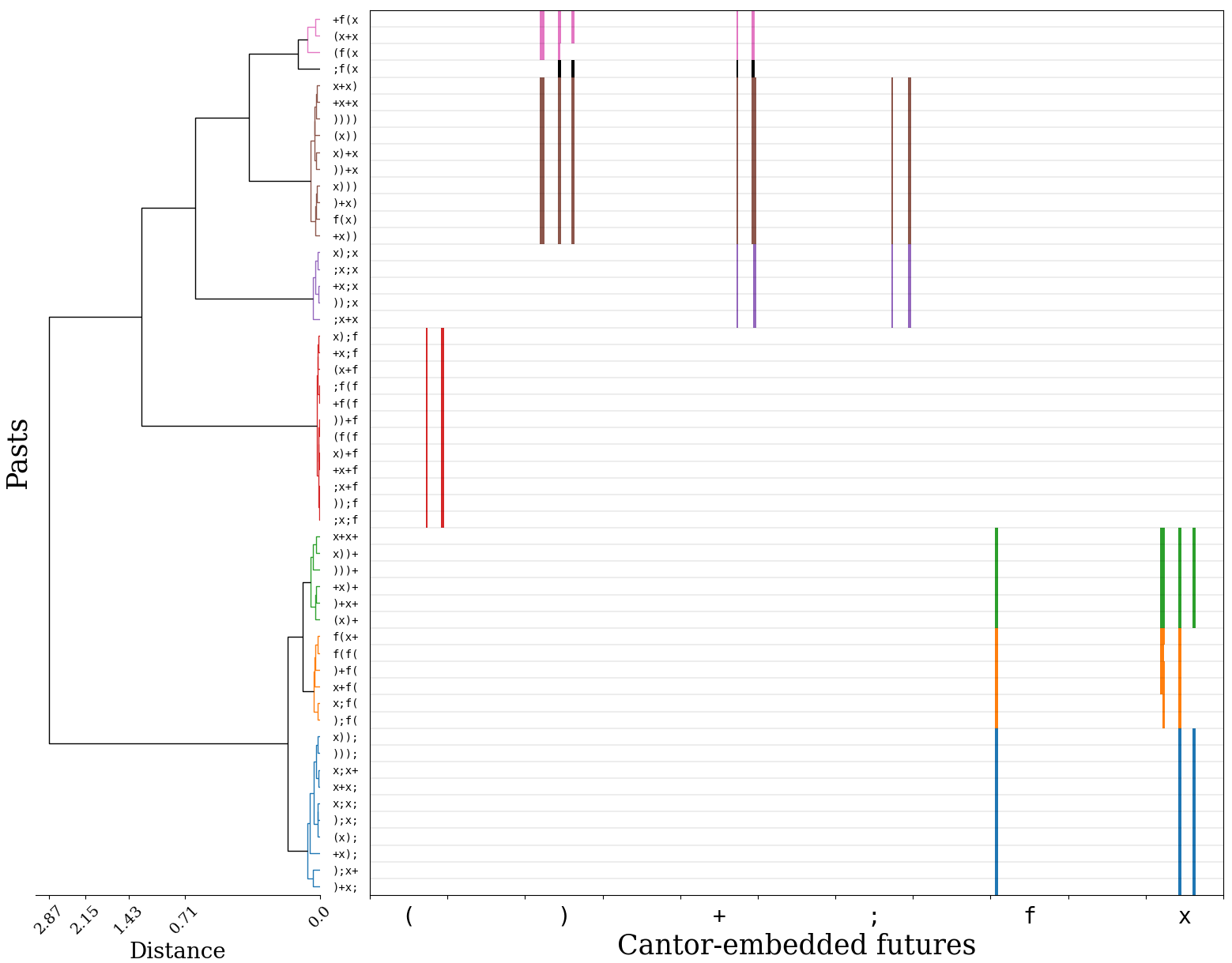}
\caption{(Upper left to lower right) Clustered Cantor diagrams of the even,
	$\mathtt{a}^n\mathtt{b}^n$, $\mathtt{a}^n\mathtt{b}^n\mathtt{c}^n$, and
	$\mathtt{x+f(x)}$ processes. Zoom for detail. For each, the vertical axis
	shows all pasts of a given length $k$ along with their hierarchically
	clustered dendrogram. $k=8$ for the even, $\mathtt{a}^n\mathtt{b}^n$, and
	$\mathtt{a}^n\mathtt{b}^n\mathtt{c}^n$ processes and $k=4$ for the
	$\mathtt{x+f(x)}$ process. For present purposes, the coloring threshold
	was chosen to aid visual interpretation. The lines in each row show the
	empirical distribution of Cantor-embedded futures observed following each
	past. As such, the horizontal axis corresponds exactly to the vertical axis
	of Fig. \ref{fig:cantor}.
	}
\label{fig:hclust}
\end{figure*}

\section{Interpretable predictive states with hierarchical clustering}

Figure \ref{fig:hclust} displays the result of collecting the Cantor-embedded
empirical predictions for all pasts of a given length for four processes---even,
$\mathtt{a}^n \mathtt{b}^n$, $\mathtt{a}^n \mathtt{b}^n \mathtt{c}^n$, and
$\mathtt{x+f(x)}$. For each, the Wasserstein distance between every pair of
predictions was computed and used to hierarchically cluster the pasts with
others that produced similar predictions, using the Ward method \cite{Mull11a}.

The resulting clustered Cantor plots offer a highly interpretable visualization
of the relationship between pasts and futures and of the predictive states'
geometry. Each plot, in a certain sense, sorts the columns in the Cantor
fractals of Fig. \ref{fig:cantor} with the white space between columns removed.
For instance, the even process's clustered Cantor plot clearly contains the two
major states, with a third ``transient'' state visible. (The latter corresponds
to the increasingly unlikely event of never seeing a $0$ in a block of length
$n$.) This third state was previously hidden mostly out of view on the
far-right side of the $2$-dimensional Cantor plot of the even process in Fig.
\ref{fig:cantor}.

Other features are worth calling out. Close observation shows that hierarchical
clustering reveals the (mostly) scale-free distinctions between pasts with
subtle differences. For the $\mathtt{a}^n \mathtt{b}^n$ process, pasts of the
form $\dots \mathtt{b a}^n$ are distinguished for different $n$, as each
involves a distinct number of $\mathtt{b}$'s appearing in the near future.
Meanwhile, the clustering scheme carefully distinguishes pasts of the form
$\dots \mathtt{b a}^n \mathtt{b}^{n-k}$ for different $k$ but \emph{not} for
different $n$, as $k$ is the essential variable for predicting the remaining
number of $\mathtt{b}$'s. (The scale-free discernment of the algorithm breaks
down past $n=5$---the scale at which sampling error becomes relevant for our
chosen sample size.)

Similar discernment is seen for the $\mathtt{a}^n \mathtt{b}^n \mathtt{c}^n$
and $\mathtt{x+f(x)}$ processes as well. We draw attention to the manner in
which the presence of a semicolon in pasts from the $\mathtt{x+f(x)}$ process
affects the comparison of predictions.

By analyzing clustered Cantor plots, one gains insight into the properties of
pasts that make them similar in terms of future predictions, even if they are
superficially quite distinct. Furthermore, the horizontal axis allows for
continued use of the Cantor set's natural geometry for visualizing the future
forecasts associated with each cluster of predictions.

\begin{figure*}[ht]
\centering
\includegraphics[width=0.49\linewidth]{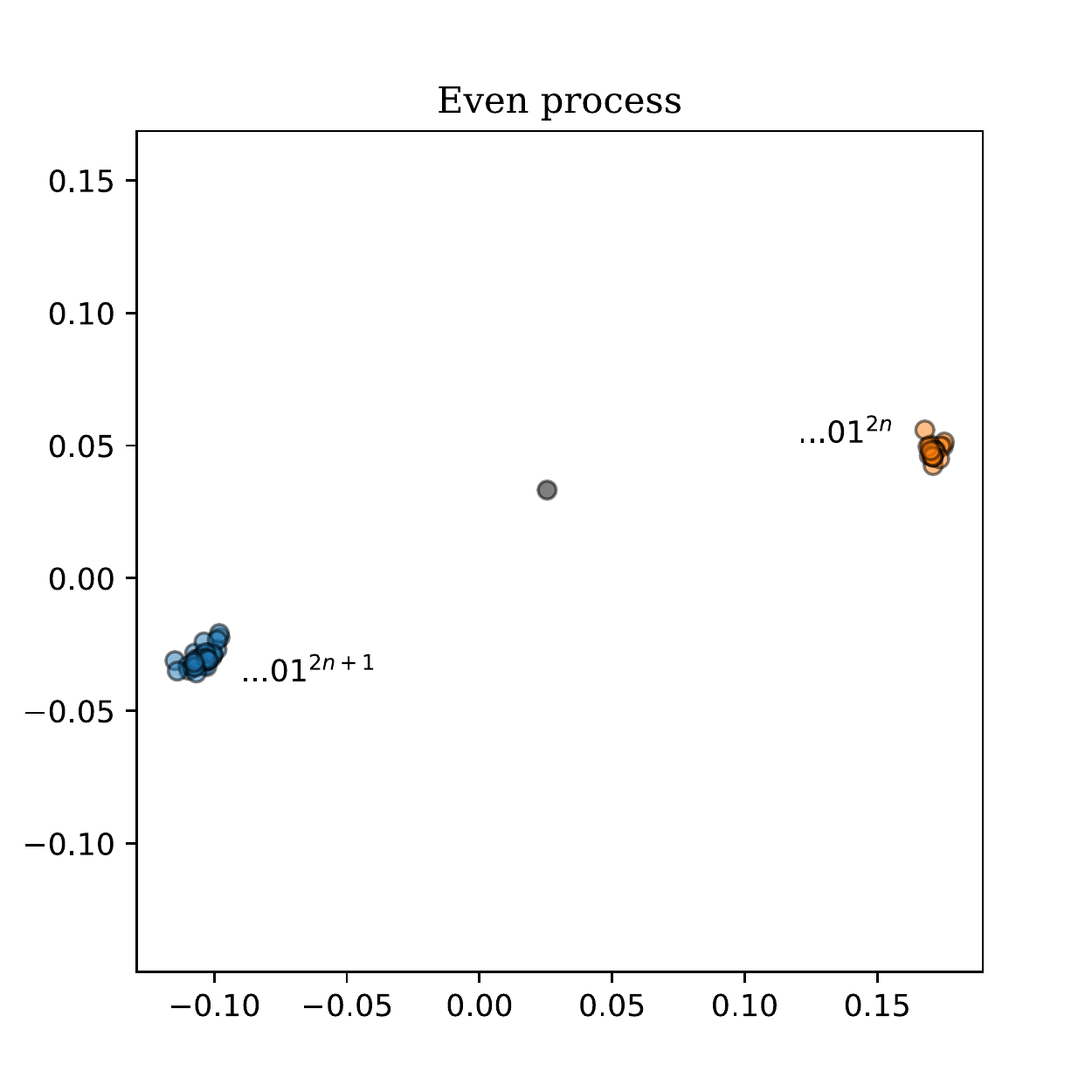}
\includegraphics[width=0.48\linewidth]{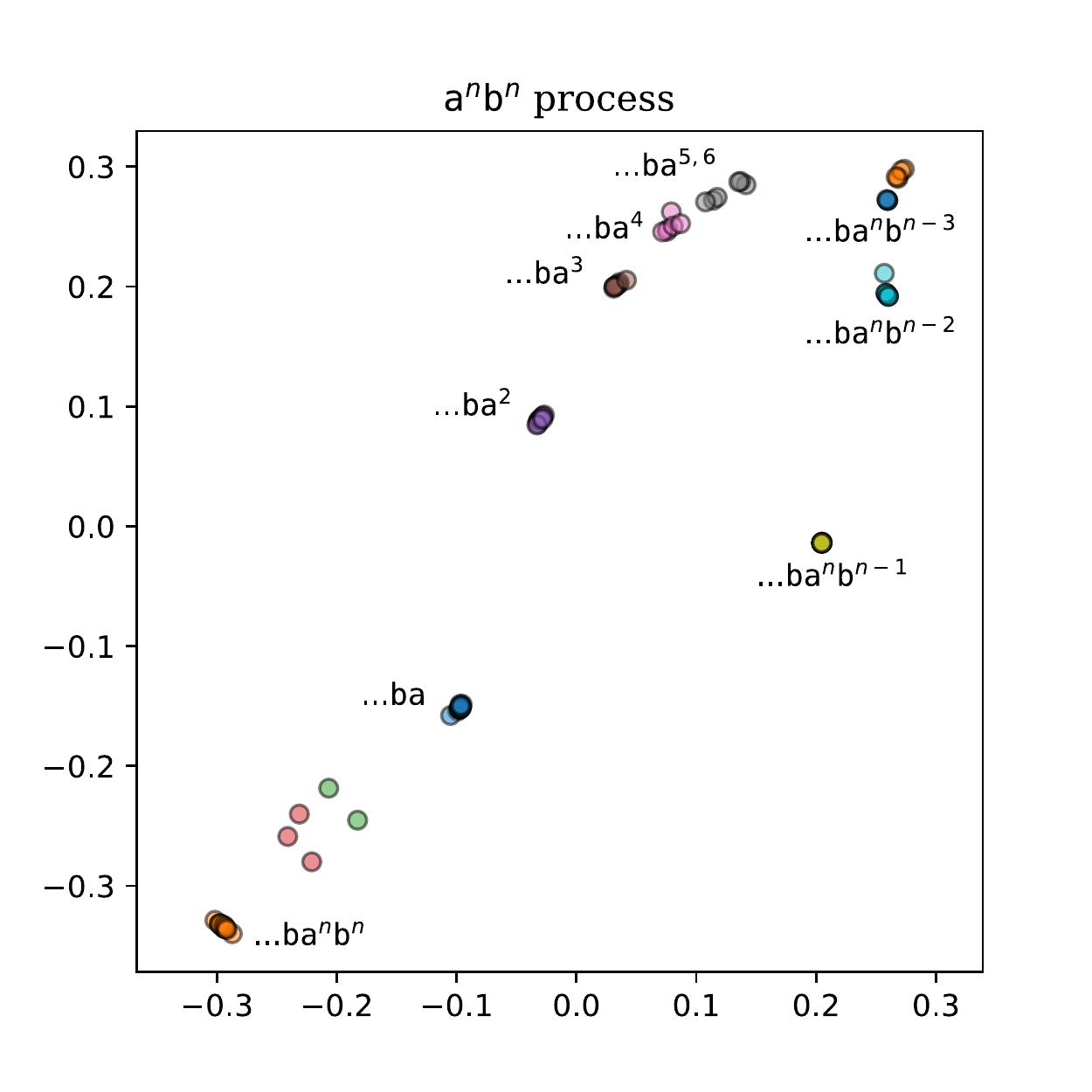}
\caption{Scatterplots of the first two MDS coordinates of the reconstructed
	predictive states: (Left) Even process. (Right) $\mathtt{a}^n\mathtt{b}^n$
	process. Clusters colored according to the scheme determined by the
	dendrogram in Fig. \ref{fig:hclust} and the label on each cluster describes
	the pattern that uniquely characterizes the pasts in that cluster.
	}
\label{fig:mds}
\end{figure*}

\section{Predictive state geometry with multidimensional scaling}

Sacrificing direct visualization of future predictions leads to a more
intuitive picture of predictive-state space geometry. Applying any desired
dimension reduction algorithm to the matrix of Wasserstein distances between
predictions yields a coordinate representation of the similarities between
predictive states.

Figure \ref{fig:mds} plots the first two dimensions of a multidimensional
scaling (MDS) decomposition \cite{Borg05a} for the even and
$\mathtt{a}^n\mathtt{b}^n$ processes. Clusters are colored in the same manner
as in Fig. \ref{fig:hclust} and labeled by the specific pattern that
distinguishes the pasts in some of the clusters. Note that the clusters and
labels are directly drawn from Fig. \ref{fig:hclust} for reference. They are
not the result of the MDS algorithm itself. However, interactive plotting
approaches may allow for similar exploration from these decompositions without
the need for prior clustering.

The even process, as in all other cases seen thus far, has two dominant
prediction clusters. These correspond to the predictive states that result
from seeing an even-sized block of $1$s (or, equivalently, no $1$s) and that
result from seeing an odd-sized block of $1$s. The lone cluster in the middle
corresponds to a transient state induced by seeing all $1$s and then a $0$.
The latter then synchronizes to right cluster.

The $\mathtt{a}^n\mathtt{b}^n$ plot is much more sophisticated. Intriguingly,
its geometry not only clearly distinguishes predictively distinct states, but
organizes them in a manner highly suggestive of an \emph{pushdown stack}. The
latter is particularly appropriate given that stack automata are the natural
analog of hidden Markov chains but for context-free languages. Observing more
$\mathtt{a}$s pushes more symbols onto the stack, with the predictive states
moving further up towards the plot's upper-right corner. And, as more
$\mathtt{b}$'s are observed the top symbol is popped off the stack, and the
predictive states move back towards the lower left. The latter represents
equality between $\mathtt{a}$s and $\mathtt{b}$s.

\begin{figure*}[ht]
\centering
\includegraphics[width=\linewidth]{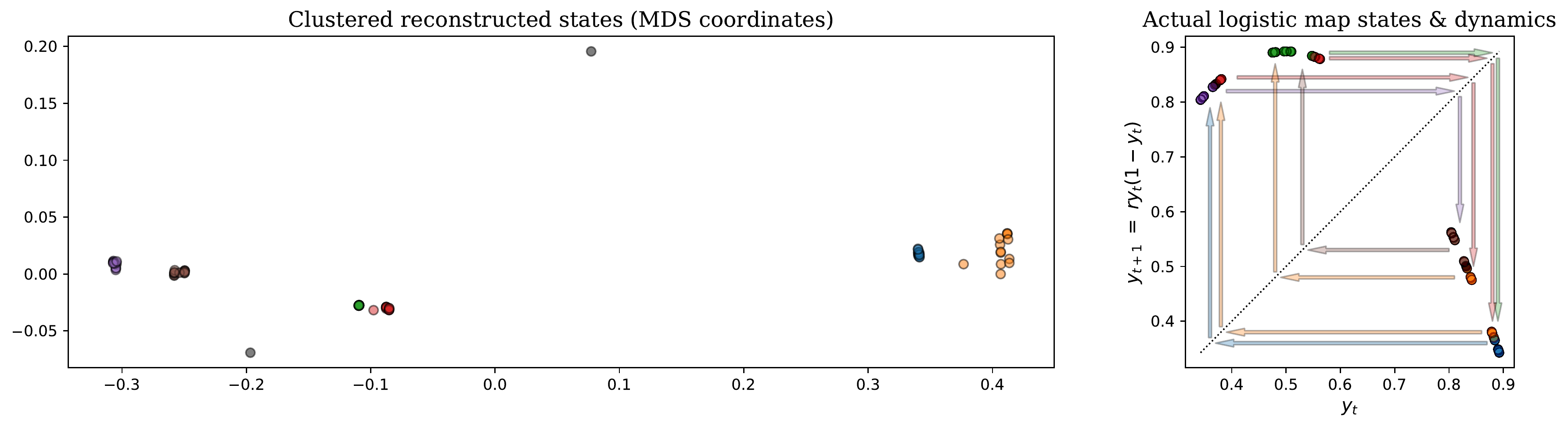}
\caption{(Left) Scatterplot of the first two MDS coordinates of the
	reconstructed predictive states for the Morse-Thue process, color-coded by
	cluster. (Right) Scatterplot of the corresponding points in the domain of
	the logistic map, plotting for each point both the present value $y_t$ and
	the next value $y_{t+1}$, with the $x=y$ line for reference. Each pair of
	color-coded arrows shows where each cluster maps to under the action of the
	logistic map. The predictively reconstructed clusters thus correspond to
	dynamically similar neighborhoods of the logistic map domain.
	}
\label{fig:mds-feig}
\end{figure*}

The geometric approach is particularly insightful when computing the
Wasserstein matrix between predictions estimated from Morse-Thue process data.
Recall that the Morse-Thue process is a coarse-graining of the iterated
logistic map $y_{t+1} = ry_t(1-y_t), ~y_0 = 0,$ at the critical chaos parameter $r_c
\approx 3.56995$. The resulting stream of $0$s and $1$s is a well-known
instance of high complexity at the ``order-disorder border''. Specifically,
setting parameter $r$ on either side of $r_c$ results in sequences that
can be generated by finite hidden Markov chains. However, at $r_c$ itself the
resulting Morse-Thue process is context-sensitive and therefore requires
infinite predictive states. That is, when it comes to capturing its behavior,
the process is several orders higher in model complexity. It is further up the
Chomsky language hierarchy.

Despite this high order of structural complexity, the predictive state geometry
reconstructed from a sufficiently large sample of the Morse-Thue process
recovers the neighborhoods of $[0,1]$ that are relevant to the dynamics of the
original logistic map. Said differently, there is a correspondence between each
past $x_{-k+1}\dots x_{0}$ and a subset $V_{x_{-k+1}\dots x_{0}}$, such that
$V_{x_{-k+1}\dots x_{0}}$ is the set of all points $y$ for which $x(f^{-t}(y)) =
x_{-t}$ for $0\leq t < n$. (Here, $f(y) = ry(1-y)$ and $x(y)$ is the encoding
$y\mapsto \{0,1\}$.) As it happens, pasts $x_{-k+1}\dots x_{0}$ whose predictive
states are close under the Wasserstein distance are also pasts for which the
sets $f(V_{x_{-k+1}\dots x_{0}})$ are close. That is, they correspond to
predictively similar ranges of the logistic map variable. 

Figure \ref{fig:mds-feig} directly visualizes the relationship between the
reconstructed predictive states of the Morse-Thue process, neighborhoods of the
logistic variable $y$, and the logistic map dynamics. In short, despite the
fact that the Morse-Thue process is a highly coarse-grained form of the
logistic map, the essential geometry of that map can be recovered by
reconstructing predictive state geometry with the Wasserstein metric and the
Cantor embedding.

Note that, due to the deterministic nature of the Morse-Thue process, the
combination of the Wasserstein metric and the Cantor embedding is particularly
important to achieving this result. Asymptotically, each past corresponds to a
unique future. And so, there is asymptotically no overlap between predictions.
The choice of the Cantor map facilitates placing together forecasts that match
up to a certain time in the future. And, the Wasserstein distance allows
directly comparing predictions whose supports are geometrically close. In this
way, the combination of the two approaches enables the straightforward recovery
of the underlying dynamical system's (logistic map's) geometry.

\section{Concluding remarks}

We presented a general approach for predictive state analysis---Cantor fractal
embedding sequences and Wasserstein distance comparison of predictions. We
offered two approaches to visualizing the results of this method---one a direct
application of multidimensional scaling and the other being a clustered Cantor
diagram built from combining hierarchical clustering with the introduced Cantor
embedding.

Compared to using reproducing kernel Hilbert spaces---a dominant approach to
predictive states at present
\cite{Song09a,Song10a,Boot13a,Brod20a,Loom21a}---our combining the Cantor set
with the Wasserstein distance may appear idiosyncratic. However, as the results
demonstrated, there are strong benefits to both and together the two methods
synergize their benefits in a unique way. The topology of convergence in
distribution can be replicated with both the Wasserstein distance and the RKHS
inner product. However, the Wasserstein distance depends on far fewer
parameters---such as, the choice of the eponymous kernel in RKHS approaches.
Moreover, its value is directly interpretable in terms of the shapes of the
distributions it compares.

Similarly, there are many ways to metrize the product topology on sequences, but
the Cantor embedding offers a direct way to connect the product topology with a
visualizable geometry. And, embedding in a single dimension enables efficient
computation of the Wasserstein metric. The benefits of the Cantor and
Wasserstein approaches adds interpretability to the resulting predictive-state
geometry along two distinct axes, most clearly seen in Fig. \ref{fig:hclust}'s
clustered Cantor diagrams. We hope that the success of this approach in
providing clear insights will complement existing thrusts in the direction of
abstract embeddings and mathematical formalism by motivating further development
on interpretable approaches to predictive state analysis.

\vspace{-0.2in}
\section*{Acknowledgments}
\vspace{-0.2in}
We thank Nicolas Brodu, Adam Rupe, Alex Jurgens, David Gier, and Mikhael
Semaan. JPC acknowledges the kind hospitality of the Telluride Science
Research Center, Santa Fe Institute, Institute for Advanced Study at the
University of Amsterdam, and California Institute of Technology for their
hospitality during visits. This material is based upon work supported by, or
in part by, Templeton World Charity Foundation Diverse Intelligences grants TWCF0440 and TWCF0570 and Foundational Questions Institute and Fetzer Franklin Fund grant number FQXI-RFP-CPW-2007 and grants W911NF-18-1-0028 and
W911NF-21-1-0048 from the U.S. Army Research Laboratory and the U.S. Army
Research Office.  

\vspace{-0.2in}
\section*{Reproducibility statement}
\vspace{-0.2in}
For the purposes of reproducibility, we provide a 
GitHub repository \cite{repo} containing
the code necessary to generate this manuscript and its figures, including a
notebook for generating the data used for the examples and for building both
static and interactive figures for further exploration.


\end{document}